\documentclass{emulateapj}
\usepackage{natbib}
\usepackage{graphicx}
\slugcomment{ApJL, in prep.}
\shorttitle{An Extremely Isolated Stellar System}
\shortauthors{Sand et al.}

\begin{document}
 \title{{\it Hubble Space Telescope} Imaging of the Ultra-Compact High Velocity Cloud AGC~226067: A stripped remnant in the Virgo Cluster}
% \title{A Hubble Space Telescope Study of AGC~226067: Isolated, Intracluster star formation in the Virgo Cluster}

\author{D. J. Sand,$\!$\altaffilmark{1} A. C. Seth,$\!$\altaffilmark{2} D. Crnojevi\'{c},$\!$\altaffilmark{1} K. Spekkens,$\!$\altaffilmark{3} J. Strader,$\!$\altaffilmark{4} E.~A.~K. Adams,$\!$\altaffilmark{5} N. Caldwell,$\!$\altaffilmark{6} P.~Guhathakurta,\altaffilmark{7} J. Kenney,\altaffilmark{8} S. Randall,\altaffilmark{6} J.~D.~Simon,\altaffilmark{9} E.~Toloba,\altaffilmark{10} B. Willman$\!$\altaffilmark{11,12}} \email{david.sand@ttu.edu}

\begin{abstract}
We analyze the optical counterpart to the ultra-compact high velocity cloud AGC~226067, utilizing imaging taken with the Advanced Camera for Surveys (ACS) on the {\it Hubble Space Telescope}.  %The color magnitude diagram of the main body of AGC~226067 reveals an exclusively young stellar population, with age $\sim$7--50 Myr and [Fe/H]$\sim$$-$0.3, and is consistent with a distance of $D$$\approx$17 Mpc, suggesting an association with the Virgo cluster.  
The color magnitude diagram of the main body of AGC~226067 reveals an exclusively young stellar population, with an age of $\sim$7--50 Myr, and is consistent with  a metallicity of [Fe/H]$\sim$$-$0.3 as previous work has measured via HII region spectroscopy.  Additionally, the color magnitude diagram is consistent with a distance of $D$$\approx$17 Mpc, suggesting an association with the Virgo cluster.  
A secondary stellar system located $\sim$1.6' ($\sim$8 kpc) away in projection has a similar stellar population.  The lack of an old red giant branch ($\gtrsim$5 Gyr) is contrasted with a serendipitously discovered Virgo dwarf in the ACS field of view (Dw J122147+132853), and the total diffuse light from AGC~226067 is consistent with the luminosity function of the resolved $\sim$7--50 Myr stellar population.  The main body of AGC~226067 has a $M_{V}$=$-$11.3$\pm$0.3, or $M_{stars}$=5.4$\pm$1.3$\times$10$^4$ $M_{\odot}$ given the stellar population.
We searched 20 deg$^2$ of imaging data adjacent to AGC~226067 in the Virgo Cluster, and found two similar stellar systems dominated by a blue stellar population, far from any massive galaxy counterpart -- if this population has similar star formation properties as AGC~226067, it implies $\sim$0.1 $M_{\odot}$ yr$^{-1}$ in Virgo intracluster star formation.  Given its unusual stellar population, AGC~226067 is likely a stripped remnant and is plausibly the result of compressed gas from the ram pressure stripped M86 subgroup ($\sim$350 kpc away in projection) as it falls into the Virgo Cluster. %, a process that recent simulations have predicted.  %Further multi-wavelength investigation of AGC~226067 and our other examples of similar stellar systems will elucidate their nature.

%While AGC~226067 may be a dwarf galaxy system which has only recently formed stars, it is most likely the result of compressed gas from the ram pressure stripped M86 subgroup ($\sim$350 kpc away in projection) as it falls into the Virgo Cluster, a process that recent simulations have predicted.  Further multi-wavelength investigation of AGC~226067 and our other examples of similar stellar systems will elucidate their nature.

\end{abstract}
 
%\altaffiltext{*}{}
\altaffiltext{1}{Texas Tech University, Physics \& Astronomy Department, Box 41051, Lubbock, TX 79409-1051, USA}
\altaffiltext{2}{Department of Physics and Astronomy, University of Utah, Salt Lake City, UT 84112, USA}
\altaffiltext{3}{Royal Military College of Canada, Department of Physics, PO Box 17000, Station Forces, Kingston, Ontario, Canada K7K 7B4}
\altaffiltext{4}{Center for Data Intensive and Time Domain Astronomy, Department of Physics and Astronomy, Michigan State University, 567 Wilson Road, East Lansing, MI 48824, USA}
\altaffiltext{5}{ASTRON, Netherlands Institute for Radio Astronomy, Postbus 2, 7900 AA Dwingeloo, The Netherlands}
\altaffiltext{6}{Harvard-Smithsonian Center for Astrophysics, Cambridge, MA 02138, USA}
\altaffiltext{7}{UCO/Lick Observatory, University of California, Santa Cruz, 1156 High Street, Santa Cruz, CA 95064, USA}
\altaffiltext{8}{Yale University Astronomy Department, P.O. Box 208101, New Haven, CT 06520-8101, USA}
\altaffiltext{9}{Observatories of the Carnegie Institution for Science, 813 Santa Barbara Street, Pasadena, CA 91101, USA}
\altaffiltext{10}{Department of Physics, University of the Pacific, 3601 Pacific Avenue, Stockton, CA 95211, USA}
\altaffiltext{11}{Steward Observatory, University of Arizona, 933 North Cherry Avenue, Tucson, AZ 85721, USA}
\altaffiltext{12}{LSST, University of Arizona, 933 North Cherry Avenue, Tucson, AZ 85721, USA}
\section{Introduction}

%Faint dwarf galaxies both in the Local Group and beyond provide constraints on cosmology as well as the interplay between baryonic physics and dark matter \citep[e.g.][]{}.  Another sentence here.  Given this, searching for faint dwarf galaxies in novel environments may provide hints as to the dominant mechanisms that drive dwarf galaxy formation and evolution.

Searching for faint, isolated dwarf galaxies in large area HI surveys has a long history \citep[see][for a review]{HIreview}.
Most recently, so-called Ultra-Compact High Velocity Clouds (UCHVCs) of neutral hydrogen have been identified by the Galactic Arecibo $L$-Band Feed Array HI (GALFA-HI) and Arecibo Legacy Fast Arecibo $L$-Band Feed Array (ALFALFA) surveys \citep[][respectively]{Saul12,Adams13} as potential sites of gas-bearing dark matter halos.  Because the spatial distribution and physical properties of these UCHVCs are consistent with some being faint dwarf galaxies in the Local Volume \citep[e.g.][]{Giovanelli10,Adams13}, several searches have been undertaken to characterize their optical properties \citep{Bellazzini15,Tollerud15,Sand15,Janesh15}.  These searches have yielded only a handful of plausible dwarf galaxy counterparts to the UCHVCs, while followup HI observations of others suggest that some may be `Dark' systems \citep[e.g.][]{Adams16}. %which never formed a significant stellar population \citep{Davies06}; 
Most, however, are likely Galactic HI clouds.

\begin{figure*}
\begin{center}
%\mbox{ \epsfysize=7.0cm \epsfbox{HST_UCHVC.eps}} 
\mbox{ \epsfysize=12.0cm \epsfbox{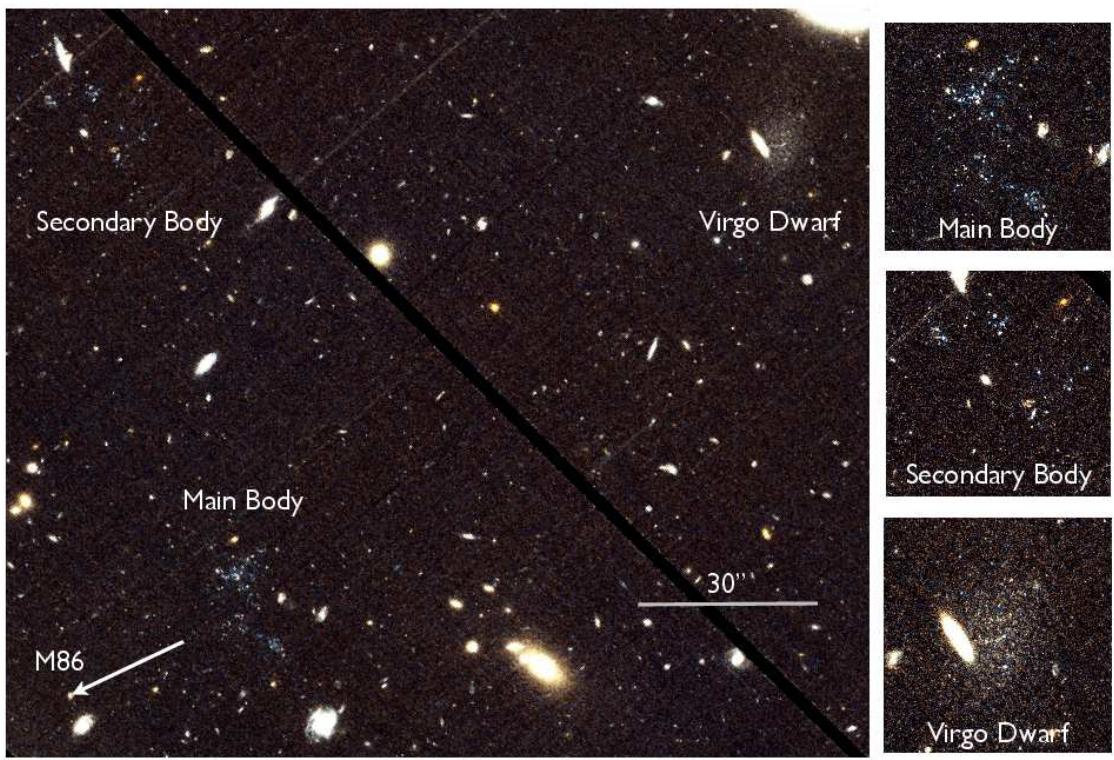}} 
\caption{     {\bf Left} A colorized cutout of the {\it F606W} and {\it F814W} {\it HST}/ACS image of  AGC~226067 and its surroundings; several notable regions are highlighted. North is up and East is to the left.  The M86 subgroup is in the Southwest direction.  {\bf Right Column} From top to bottom we show zoom-ins of the main body of AGC~226067, its secondary body and a newly discovered Virgo dwarf galaxy,  Dw~J122147+132853.   \label{fig:image}}
\end{center}
\end{figure*}

The subject of this paper is {\it Hubble Space Telescope} ({\it HST}) imaging of AGC~226067 ($v_{\odot}$=$-$140 km s$^{-1}$; also referred to as SECCO1), a UCHVC with an intriguing optical counterpart.  First identified by \citet{Bellazzini15}, it was confirmed spectroscopically to be associated with the coincident HI cloud \citep[][]{Bellazzini15UCHVC,Sand15}.    Follow-up high-resolution HI observations with the  Very Large Array (VLA) showed that  AGC~226067 broke up into two distinct HI clouds \citep{Adams15} -- the primary HI source is coincident with the optical/UV counterpart, while the secondary HI component also has an optical/UV counterpart offset by 0.5'.  Data from the Multi Unit Spectroscopic Explorer (MUSE) at the VLT revealed 38 separate HII regions scattered throughout the main and secondary stellar body of AGC~226067, with a nearly uniform metallicity of $\langle$12 + log(O/H)$\rangle$ = 8.37 (for the main body), more metal-rich than expected if AGC~226067 was a gas-rich dwarf galaxy \citep{Beccari16b}.  %An initial {\it HST} analysis revealed little evidence for an underlying old stellar population.
AGC~226067 is projected onto the `Low Velocity Cloud' region of the Virgo Cluster \citep[at $D$$\approx$17Mpc, which we use throughout this work;][]{Boselli14}, which has a velocity distribution centered at $v_{LSR}$$\sim$0 km s$^{-1}$ with a range of  $\pm$400 km s$^{-1}$.  
%This environment, and the lack of an apparent old stellar population \citep[][to be discussed further here]{Beccari16b}, led to suggestions that AGC~226067 may be a blue, star-forming galaxy with especially low surface brightness \citep{Adams15} or may originate from stripped gas due to interactions in the larger Virgo environment \citep{Beccari16b}.  
For simplicity, we refer to the entire system (consisting of the main and secondary stellar bodies, along with the two HI clouds) as AGC~226067 throughout this work, and will refer to particular regions of the system (e.g. its `main' and `secondary' body) when getting into specifics.
In addition to a presentation of the {\it HST} data (Section~\ref{sec:datareduce}) and an analysis of the stellar population of AGC~226067 (Section~\ref{sec:props}), we perform an initial optical/UV archival search for similar stellar systems (Section~\ref{sec:search}) in deep Virgo cluster data.  We discuss possible origins for AGC~226067 in Section~\ref{sec:discuss}, and then conclude (Section~\ref{sec:conclude}).

%[][similar to the SHIELD galaxy sample; \citealt{Cannon11}]
%one of the few UCHVC systems with an optical counterpart.  The UCHVC neutral gas properties  An optical counterpart to AGCblahb<lah was first identified by \citet{Bellazzini14} in a Large Binocular Telescope search, and follow up spectroscopy indicated \citep{Bellazzini_UCHVC}.  These results were confirmed by \citet{Sand15a} in a comprehensive archival search AGCblahblah is projected onto the `Low Velocity Cloud' region of the Virgo Cluster \citep[at D=17Mpc;][]{Gavazzi99}, which has a velocity distribution centered on $v_{LSR}$$\sim$0 km s$^{-1}$ with a dispersion of $\sim$200 km s$^{-1}$.

\begin{figure*}
\begin{center}
\mbox{ \epsfysize=8.0cm \epsfbox{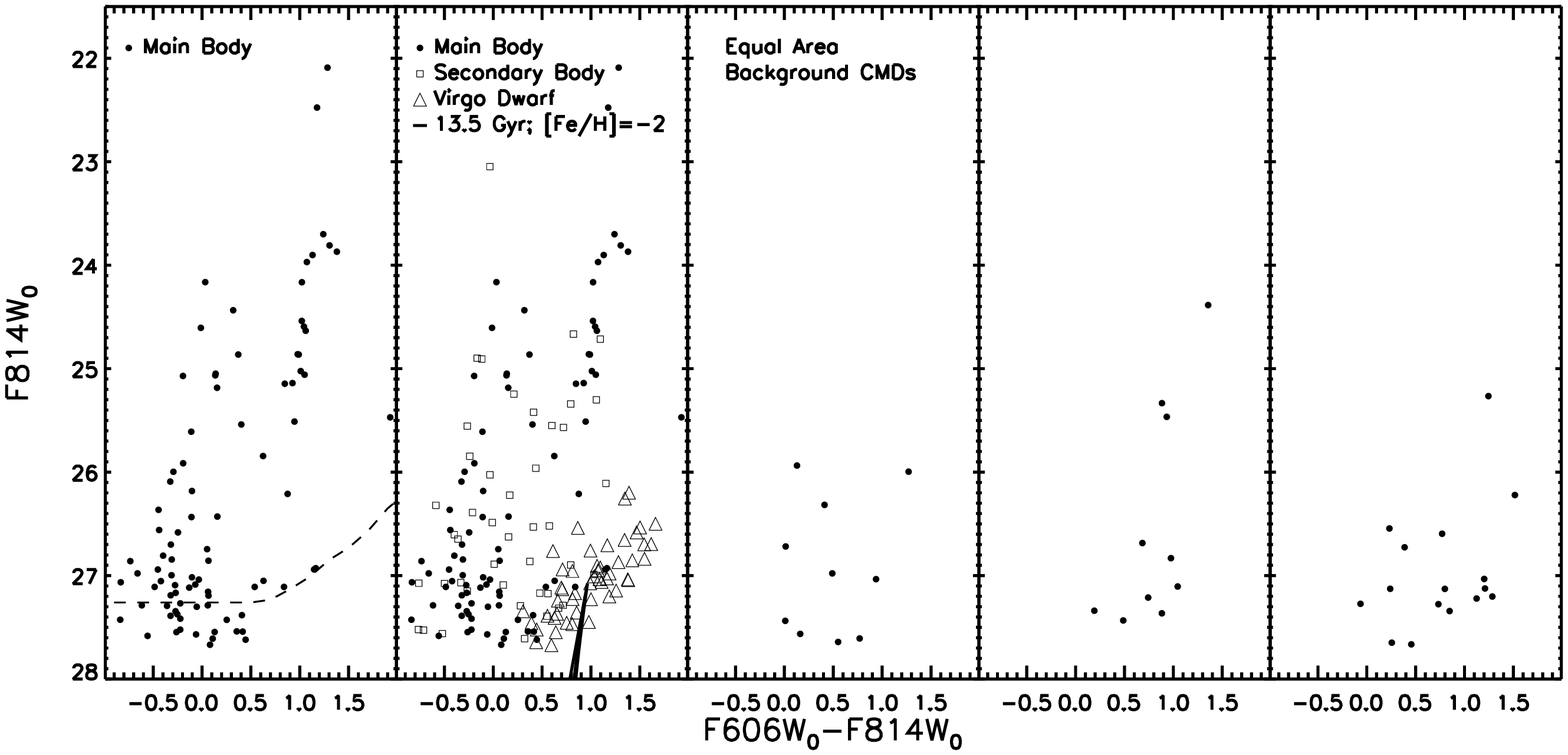}} 
\mbox{ \epsfysize=8.0cm \epsfbox{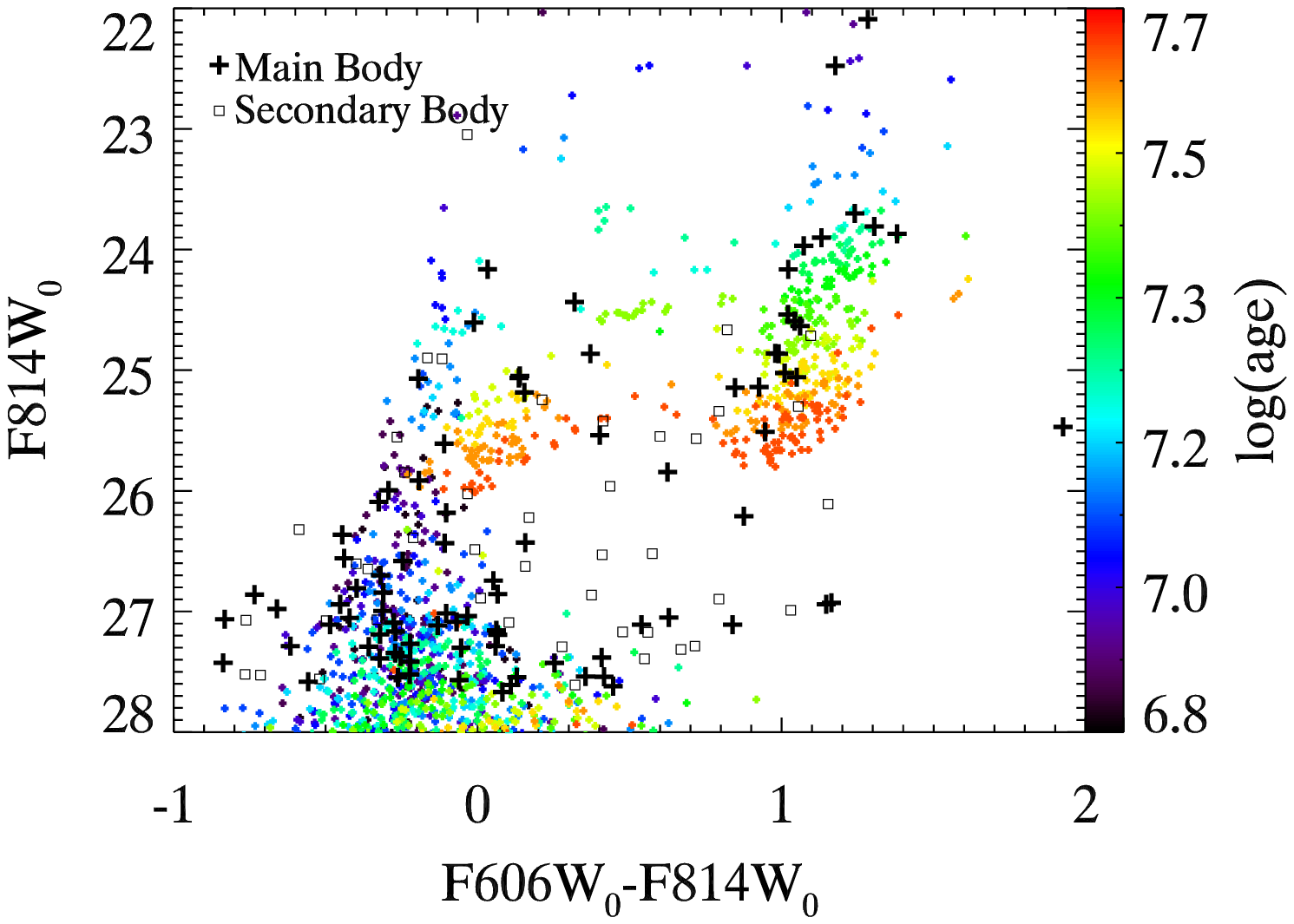}} 
%\mbox{ \epsfysize=8.0cm \epsfbox{../photom/ALFALFA_cmd_otherdw.eps}} 
%\mbox{ \epsfysize=8.0cm \epsfbox{../photom/ALFALFA_cmd_sec.eps}} 
\caption{  {\bf Top:} The left panel shows the CMD of the main body of AGC~226067,  followed by the CMDs of the secondary body and Dw~J122147+132853.  While the secondary body has a CMD similar to that of the main body, Dw~J122147+132853 has a clear RGB population that is not apparent in AGC~226067.  The three adjacent panels show equal area background CMDs.  The CMD of AGC~226067 shows an overdensity of blue stars and a RHeB sequence.  The dashed line in the far left panel shows the 50\% completeness limit.  In Table~1 we list the position and area over which each of the non-background CMDs is drawn from.  {\bf Bottom:} The CMD of the main and secondary body, along with a simulated stellar population with a constant star formation history between $\sim$7--50 Myr and a [Fe/H]$\sim$$-$0.3, which matches the extent of the observed RHeB sequence.  %See Section~\ref{sec:stellarpops} for a discussion of its star formation history. 
 \label{fig:cmds}}
\end{center}
\end{figure*}

\section{{\it Hubble Space Telescope} Observations \& Data Reduction} \label{sec:datareduce}

%F814W date: Apr 26 2015
%F606W date: Apr 26 2015
%F275W date: Jul 7 2015

%Observations of AGC~226067 were taken with the Advance Camera for Surveys \citep[ACS;][]{ACS} in the F606W (2196 s) and F814W (2336 s) filters, as well as the Wide Field Camera 3 \citep[WFC3;][]{} in the F275W filter (2470 s) aboard HST as part of HST-GO-12445 (PI: D. Sand).

HST observations of AGC~226067 (GO 13735; PI Sand) were taken with ACS \citep{ACS} in the F606W (2196 s) and F814W (2336 s) filters on 2015 April 26, as well as the Wide Field Camera 3 (WFC3) in the F275W filter (2470 s) on 2015 July 7.  The WFC3 F275W data contained only a handful of low S/N sources, and we will not consider it further.  For the $F606W$ and $F814W$ data, multiple exposures were taken in each filter to remove cosmic rays, but we did not dither to fill in the chip gaps, as this was not necessary to fully image AGC~226067.  

Point spread function photometry was performed using the software package {\sc Dolphot} \citep{Dolphin00} on the individual charge transfer efficiency corrected ACS images (the {\sc flc} files), with input parameters similar to \citet{Williams14}.  Sources with (sharp$_{F606W}$ + sharp$_{F814W}$)$^2$$>$0.1 and (crowd$_{F606W}$ + crowd$_{F814W}$)$^2$$>$1.0 were culled, and only those remaining sources with a signal to noise ratio (SNR) greater than 4 in both the F606W and F814W bands were kept for further analysis.  Artificial star tests were also run to quantify our photometric uncertainties and completeness; the 50\% completeness limits are shown in Figure~\ref{fig:cmds}. The final photometric catalog was corrected for Galactic extinction \citep{Schlafly11} corresponding to the position of the main body of AGC~226067, with an adopted value of $E(B-V)=0.057$.  All magnitudes presented in this work have this correction applied.  All results are presented in the {\sc vegamag} system.

To create mosaics of the F606W and F814W ACS images, we used Drizzlepac v2.0 and the {\sc Astrodrizzle} routine.  In Figure~\ref{fig:image}, we present a false color RGB image of AGC~226067 and its surroundings (using the average of the F606W and F814W images as our `green' image), with zoomed in cutouts highlighting several points of interest.  The main stellar body of AGC~226067 consists primarily of blue stars in several distinct clumps.  A second clump of blue stars $\sim$1.6 arcmin to the Northwest of the main body of AGC~226067 is seen, as has been noted in previous work \citep{Sand15,Adams15,Beccari16,Beccari16b}.  In contrast, an uncatalogued Virgo cluster dwarf galaxy serendipitously in our ACS field of view (which we dub Dw~J122147+132853) consists almost entirely of red giant branch (RGB) stars, which we discuss below.

\begin{figure*}
\begin{center}
\mbox{ \epsfysize=5.5cm \epsfbox{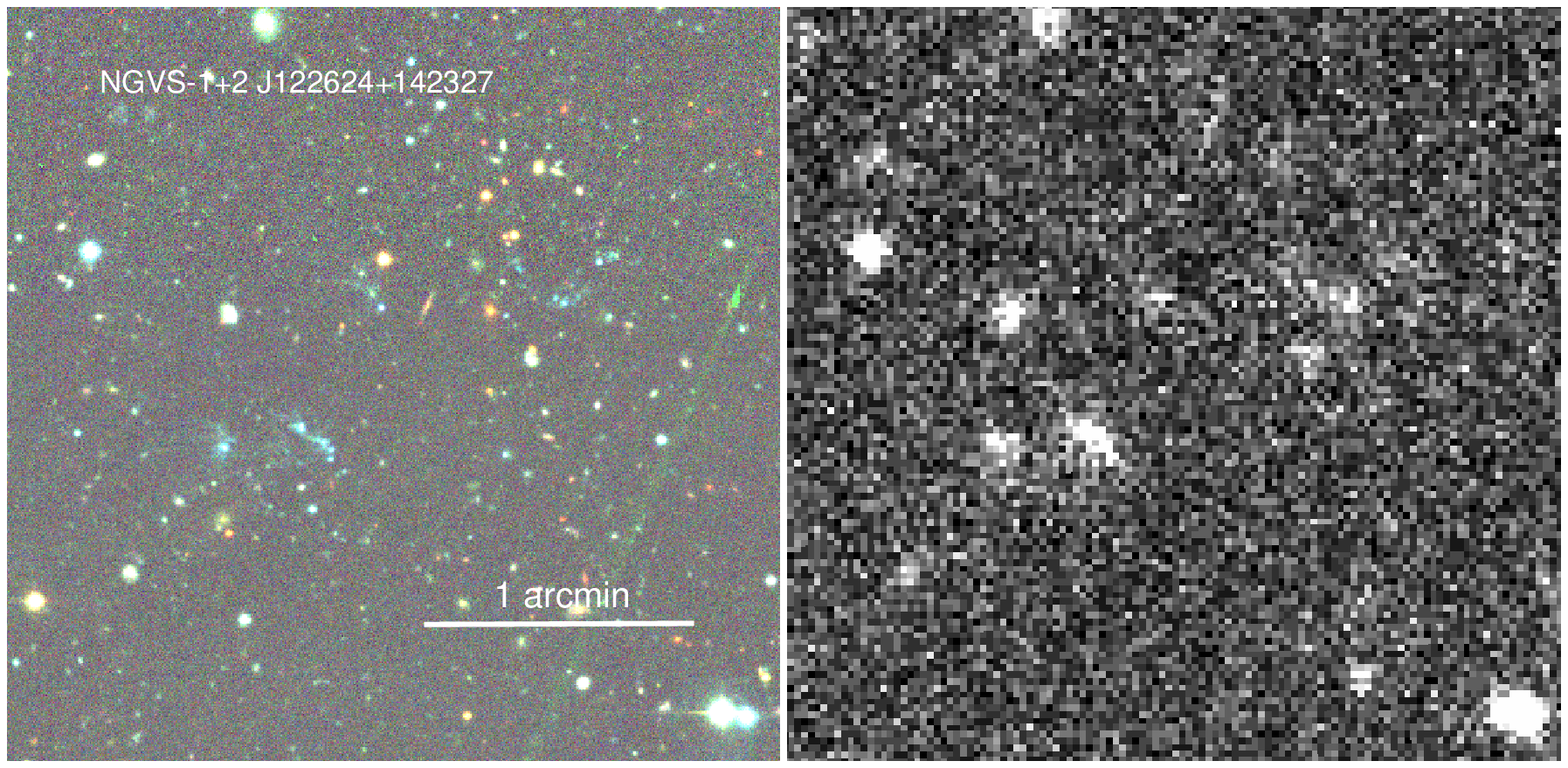}} 
\mbox{ \epsfysize=5.5cm \epsfbox{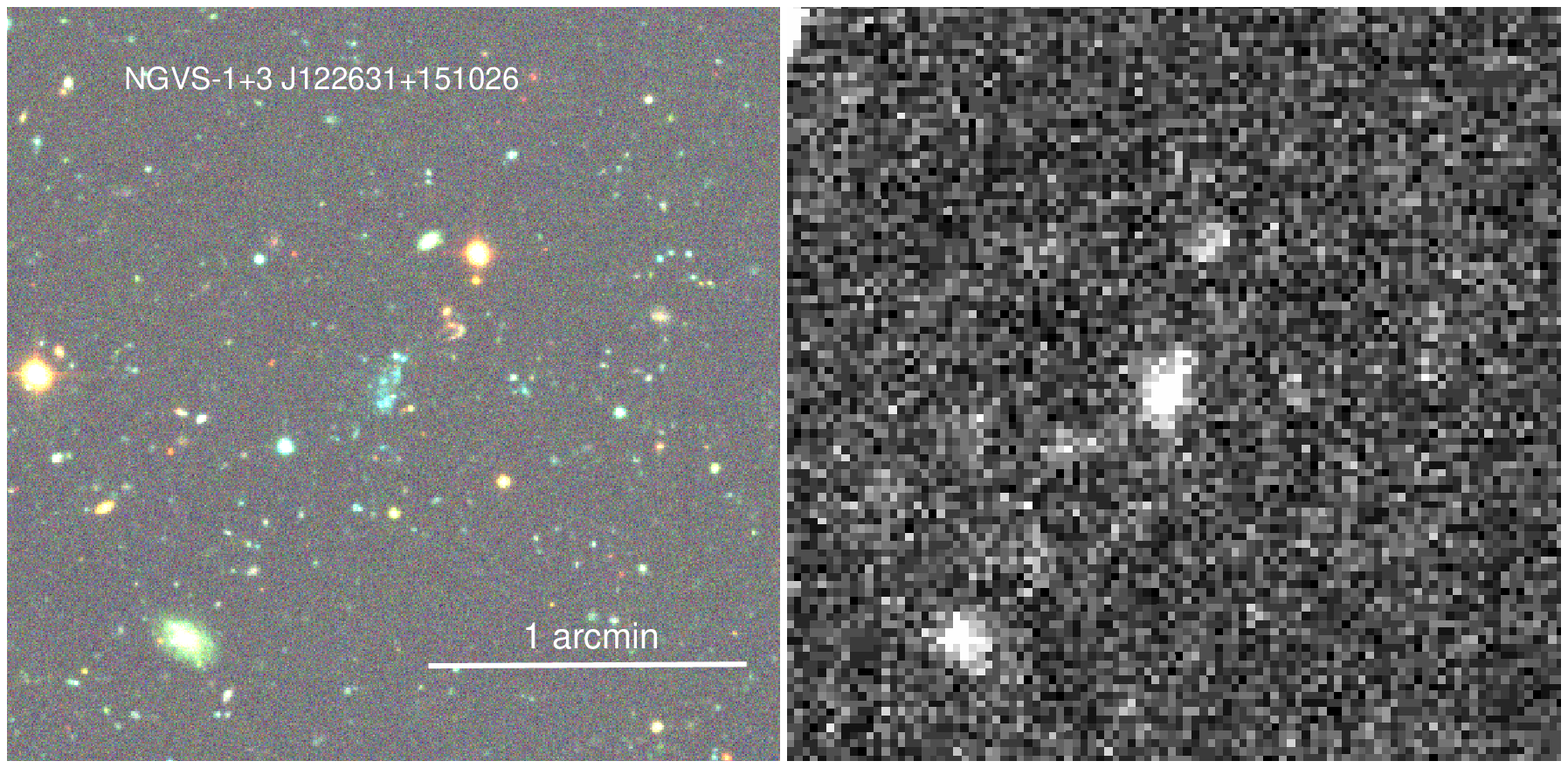}} 
\caption{  Color RGB optical (left) and {\it GALEX} NUV (right) images of the two stellar systems that look similar to AGC~226067 in a preliminary search of 20 deg$^2$ of NGVS archival data.  Both candidates project far from massive Virgo cluster galaxies and appear dominated by young stellar populations without a diffuse galaxy-like component.   North is up and East is to the left. \label{fig:newobjects}}
\end{center}
\end{figure*}

\section{Properties of AGC~226067}\label{sec:props}

\subsection{Stellar Population}\label{sec:stellarpops}

The color magnitude diagram (CMD) of AGC~226067 reveals a complex and exclusively young stellar population.  In the top panel of Figure~\ref{fig:cmds}, we show the CMD of AGC~226067 centered on its main body, along with several background CMDs drawn from equal area outlying regions.  AGC~226067's main body has a population of faint, blue stars ($F606W_0$$-$$F814W_0$$<$0.0 and $F814W_0$$\gtrsim$26 mag) that are likely young main sequence stars, along with a sequence of stars from 0.8$<$$F606W_0$$-$$F814W_0$$<$1.3 and 22$<$$F814W_0$$<$26 mag which are consistent with being red helium burning (RHeB) stars.  There are only a handful of possible old, RGB stars (at $F606W_0$$-$$F814W_0$$>$0.4 and $F814W_0$$\gtrsim$27).
The morphology of AGC226067 is broken up into several clumps with an approximate spatial extent of $\sim$20" ($\sim$1.6 kpc; Figure~\ref{fig:image}).

Also shown in Figure~\ref{fig:cmds}, we overplot the CMDs of the secondary body (to be discussed in Section~\ref{sec:secondary}) and the new Virgo dwarf Dw~J122147+132853, along with a 13.5 Gyr, [Fe/H]=$-$2 isochrone \citep{Bressan12}.  There is no apparent overdensity of RGB stars associated with AGC~226067, signaling that any old stellar population must be small.  This absence of RGB stars becomes more striking when compared with the CMD of Dw~J122147+132853, which shows a large population of such stars (compatible with our adopted $D$=17 Mpc).  A simple large-aperture analysis \citep[see][with masking of the nearby background spiral galaxy]{Sand14} indicates that Dw~J122147+132853 has an absolute magnitude of $M_{F606W}$=$-$11.3$\pm$0.2 and $M_{F814W}$=$-$11.7$\pm$0.2; this is $M_{V}$=$-$11.4$\pm$0.2 using the transformations of \citet{Sirianni05}.  %We place strict limits on an old stellar population in AGC~226067 in Section~\ref{sec:pop}.

The brightness of RHeB stars in optical CMDs is directly dependent on the age of the star \citep[][]{McQuinn11}, which we will use to constrain the stellar population in AGC~226067.  In the bottom panel of Figure~\ref{fig:cmds}, we show the results of simulating a constant star formation history 1$\times$10$^6$ $M_{\odot}$ stellar population with  ages between $\sim$7--50 Myr and a [Fe/H]=$-$0.3 \citep[corresponding to the gas-phase metallicity found by ][assuming the solar abundance of \citealt{Grevesse07}]{Beccari16b} at a distance of 17~Mpc (we assumed a Kroupa initial mass function; \citealt{Kroupa01}).  We have convolved the simulated data set with our measured photometric uncertainties.  When compared with the CMD of AGC~226067's main body, the dimmest RHeB stars correspond to an age of $\sim$50 Myr, while the two bright stars at the top of the sequence ($F814W_0$$\approx$22-22.5) correspond to RHeB stars of $\sim$7-8 Myr -- we infer an approximate stellar population with age range of $\sim$7-50 Myr for AGC~226067.  Star formation could have continued to the present day at a similar rate, but our CMDs can't constrain this due to low number statistics.  
The simulated stellar population is not a perfect match to the colors of the observed CMD, but this is likely due to known limitations of stellar evolution models at these young ages \citep[i.e. uncertainties associated with internal mixing, stellar rotation and mass loss;][]{DP02,McQuinn11}.  The age of the youngest population of stars is consistent with the observation that AGC~226067 has HII regions.

%, and our CMD-based metallicity of [Fe/H]$\sim$$-$0.3 is compatible with the mean abundance of $\sim$0.5$Z_{\odot}$ found in the HII spectroscopy of \citet{Beccari16b}.

%, and we do not try to constrain the metallicity any further than to state that the CMD is consistent with [Fe/H]=$-$0.3.

We estimate the stellar mass of the main body of AGC~226067 by directly comparing the relative number of RHeB stars to that of our simulated 1$\times$10$^6$ $M_{\odot}$ stellar population, subtracting a background derived from outlying regions of the ACS field of view.  We find $M_{stars}$= 5.4$\pm$1.3$\times$10$^4$ $M_{\odot}$ (where the error bar includes the effects of Poisson statistics and background subtraction, but not variations in stellar models or star formation histories), which is consistent with our finding from direct aperture photometry on the main body (Section~\ref{sec:pop}).  Note that the photometric completeness at the faint end of the RHeB sequence is $>$95\%, and so cannot effect our results.  This and other properties of AGC~226067 are shown in Table~\ref{table:properties}.

We can use this stellar mass to obtain an estimate of the star formation rate (SFR) of the main body; assuming uniform star formation over the past 50~Myr, we obtain a SFR of 1.1$\times$10$^{-3}$~M$_\odot$/yr. This is in general agreement with revised values of the SFR based off the MUSE spectra \citep{Beccari17_error}.

We can also look at the nature of star formation in this object by examining where it falls on a Kennicutt-Schmidt relation of SFR density ($\Sigma_{SFR}$) vs.~gas density.  Assuming our SFR estimate for the main body and a radius of 1.1~kpc (Table~\ref{table:properties}), we get a $\Sigma_{SFR}$$=$2.9$\times$10$^{-4}$~M$_\odot$/yr/kpc$^2$.  The main body has an estimated HI mass of 1.5$\times$10$^{7}$~M$_\odot$ in a deconvolved area of 3.08$\times$10$^7$ pc$^2$ \citep{Adams15}, giving a gas surface density of 0.49 M$_\odot$/pc$^2$.  These values suggest the star formation in AGC~226067 is occurring at the low end of gas surface densities in comparison to nearby galaxy disks \citep{Bigiel08}, and is slightly more efficient than that typically seen in the outer regions of nearby spirals and dwarfs \citep{Bigiel10}.

\begin{figure*}
\begin{center}
\mbox{ \epsfysize=9.0cm \epsfbox{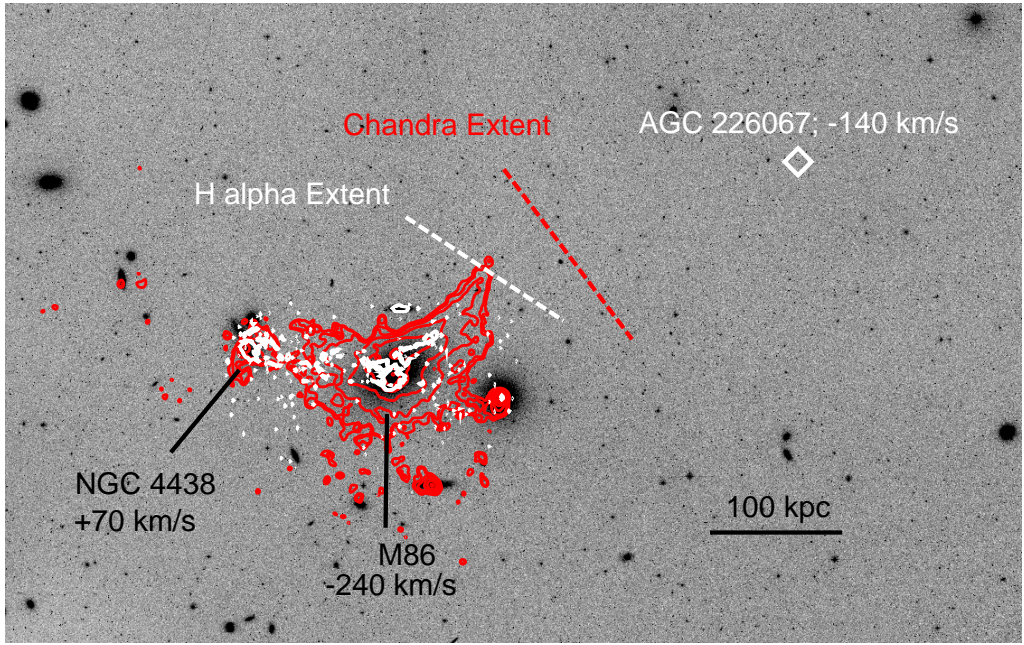}} 
\caption{  The M86 subgroup with respect to ACG~226067, which is $D$$\approx$350 kpc away in projection. We show both H$\alpha$ \citep[white;][]{Kenney08} and {\it Chandra} surface brightness contours \citep[red;][]{Randall08}, along with the rough spatial extent of each data set.  A tail of H$\alpha$ and ram pressure stripped gas -- a consequence of NGC~4438 and M86's complicated interaction with each other and the Virgo intracluster medium -- is pointing roughly in the direction of AGC~226067.  %The diamond symbol representing ACS~226067 is not representative of its actual size.
  \label{fig:environ}}
\end{center}
\end{figure*}

\subsection{The Secondary Body} \label{sec:secondary}

Several previous works have also pointed out a secondary clump of blue stars (Figure~\ref{fig:image}), located $\sim$1.6' ($\sim$8 kpc) to the Northeast of the main body of AGC~226067 \citep{Sand15,Adams15,Beccari16,Beccari16b}.  A CMD of this region is shown in the middle-left panel at the top of Figure~\ref{fig:cmds}.  The stars in this secondary knot are consistent with the stellar population of the main body of AGC~226067.  Although it is not sufficiently populated to confirm the presence of stars with age $<$10 Myr, it is likely the case given the secondary clump has several HII regions \citep{Beccari16b}.  %There are several blue stars consistent with a $\sim$7 Myr  and [Fe/H]$\sim$$-$0.6 main sequence stellar population, and several others which are consistent with the $\sim$30 Myr giant branch.  
Again, there are few old RGB stars, in stark contrast to the nearby Virgo dwarf.  
We estimate the stellar mass in the secondary body identically to the technique employed on the main body and find $M_{stars}$=5.5$\pm$4.2$\times$10$^3$ $M_{\odot}$ (see Table~1).  
%The other stars in the secondary clump CMD are consistent with being background/foreground stars as shown in the top panel of Figure~\ref{fig:cmds}.

%As the secondary clump has a similar stellar population as the main body of AGC~226067, it is likely that they are physically related, as also shown by the secondary body's HII region velocities \citep{Beccari16b}.
%although as discussed by \citet{Adams15}, there is no HI gas associated with it (unlike the main body).  
%\citet{Adams15} also identified a second clump of HI gas (AGC229490), physically distinct from both the main body of AGC~226067 and its secondary stellar clump, for which we will search for a stellar counterpart in the following section.

\subsection{Other possible stellar extensions of AGC~226067}

Two other regions around AGC~226067 have been discussed in the literature as possible physically-associated extensions, and we examine our {\it HST} photometry at these positions in more detail here.  

The first is an additional clump of HI gas, discovered and dubbed AGC~229490 by \citet{Adams15}, which is physically separated from the main body of AGC~226067, but has a consistent HI-derived velocity (v$_{\odot}$=$-$123 and $-$142 km s$^{-1}$, respectively).  We have constructed a CMD of AGC~229490 corresponding to the position of the gas cloud, and its rough dimensions \citep[RA: 12:21:53.8, DEC: +13:29:08, with size 45"$\times$45";][]{Adams15}.  The CMD is largely consistent with several equal area background CMDs, possibly with an overdensity of $\sim$5 or fewer stars.  Thus the secondary HI cloud, AGC 229490, is largely star-free.

More recently, a deep ground based image of the AGC~226067 field suggested the presence of an additional blue stellar over density located $\sim$1 arcmin to the Northeast of the main body \citep{Beccari16}.  A CMD of the region again suggests that this field consists of primarily foreground stars, although there are a handful of stars that are consistent with a $\sim$7-50 Myr stellar population.   %We do not consider this proposed overdensity further in this work.

\subsection{Constraints on an Old Stellar Population}\label{sec:pop}

The lack of RGB stars in the CMD of AGC~226067 is striking when compared with the serendipitous Virgo dwarf galaxy in our ACS field of view, which has an absolute magnitude of  $M_V$$\approx$$-$11.4.  We take two approaches for constraining an old stellar population in AGC~226067's main body.

First, we use the number of stars in the region of the CMD where we find RGB stars in Dw~J122147+132853 to constrain the old stellar mass of AGC~226067.  We defined a CMD box surrounding the stars within Dw~J122147+132853 (with $F606W_0$$-$$F814W_0$ between 0.30 and 1.66 and $F814W_0$ between 26.20 and 27.67); after background subtraction we find 43$\pm$7 in Dw~J122147+132853.  However for the main body of AGC~226067, we find only 4$\pm$4 stars.  Using the absolute magnitude of Dw~J122147+132853 ($M_V=-11.4$), this translates to a limiting absolute magnitude for the old population of AGC~226067 of $M_{V}$$>$$-$8.8 mag, or $M_{stars}$$<$6$\times$10$^5$ $M_{\odot}$ \citep[assuming a $M/L_V$=2 $M_{\odot}$/$L_{\odot}$ for an old stellar population;][]{Strader11,Baumgardt17}.  This limit on the old stellar population in AGC~226067 hinges on it being at the same distance as Dw~J122147+132853; if it were at a greater relative distance, the old stellar populations limits would be weaker.  Similarly, if any putative old stellar population in AGC~226067 were significantly more metal rich than  in Dw~J122147+132853  its RGB stars would be pushed to relatively redder colors, outside of our CMD selection box.   This relative difference in metallicity would again cause us to underestimate the old stellar population in AGC~226067.  Finally, if a population of older stars is associated with AGC 226067, but spread out over a larger area than the young stellar population, then our limits on an old stellar population would be proportionately brighter (assuming a constant stellar surface density).  

We measure the integrated magnitude for the main body within an 11'' radius around AGC~226067 (after masking background galaxies) of $M_{F606W,0}$=$-$11.3$\pm$0.3 and $M_{F814W,0}$=$-$11.2$\pm$0.3, translating to $M_{V}$=$-$11.3$\pm$0.3.  For the age range of 7-50~Myr, the \citet{BC03} models suggest $M/L_V$=0.013--0.06 $M_{\odot}$/$L_{\odot}$; this suggests a stellar mass of $M_{stars}$=4-17$\times$10$^4$ $M_{\odot}$,  consistent with our measurement of $M_{stars}$= 5.4$\pm$1.3$\times$10$^4$ $M_{\odot}$ in Section~\ref{sec:stellarpops}.  
%Thus there is no need for an underlying old population to explain the system's absolute magnitude.  
This measured absolute magnitude is fully consistent with previous ground-based data \citep[e.g.][]{Adams15}, but is $\sim$2-$\sigma$ brighter than \citet{Beccari16b}.

We also subtracted the point sources in our {\sc Dolphot} catalog from the F606W and F814W ACS mosaic images, and have analyzed the remaining diffuse light to constrain an old stellar population.  %An aperture of 220 pixels (11 arcsec) roughly centered on the main body of AGC~226067 was used,  which encompassed all of the visible diffuse light remaining in the point-source subtracted images.  We experimented with several different background annuli during this process, as well as equal-area background regions and direct contaminating galaxy removal. %in order to obtain the best results.  
Using this image and the same 11'' annulus, we measure a diffuse, star-subtracted magnitude of $M_{F606W,0}$=$-$9.8$\pm$0.4 and $M_{F814W,0}$=$-$10.0$\pm$0.5, which corresponds to $M_{V}$=$-$9.8 $\pm$0.5 mag.  The errors on these measurements include the range of values obtained using direct galaxy removal and background annulus subtraction.  

We can compare the amount of diffuse light to that expected from the faint unresolved stars of the $\sim$7-50 Myr stellar population revealed in the CMD, but below the detection limit of the ACS data.  To estimate this, we first calculated the total magnitude of resolved stars consistent with this population in the main body of AGC~226067 and used it as an anchor for integrating the luminosity function below our ACS detection limit, assuming a 7--50 Myr stellar population with [Fe/H]=$-$0.3 and the associated Padova luminosity function \citep{Bressan12} and a Chabrier initial mass function.  Depending on the choice of age of the stellar population, and the filter, our resolved stellar sources should make up $\sim$45-70\% of the total luminosity of AGC~226067, implying a diffuse stellar component from the young stellar population with %$M_{F814W,0}$=$-$10.0$\pm$0.1 and 
$M_{V}$=$-$10.4$\pm$0.2 mag (where the error bars encapsulate the allowed range when varying the age of the stellar population).  Thus, the observed diffuse component of AGC~226067 (directly measured as $M_{V}$=$-$9.8 $\pm$0.5 mag) is fully consistent with a stellar population belonging solely to the apparent young stellar population, with no need to invoke any associated older population. %A solely young stellar population, if true, would make AGC~226067 a unique dwarf galaxy if that were its origin.

Given the apparent lack of an old stellar population, AGC~226067 cannot be considered a typical dwarf irregular-like galaxy as these systems all have an old RGB population at {\it HST} depths \citep[e.g.][]{McQuinn14}.  This strongly suggests that AGC~226067 is analogous to so-called `tidal dwarfs' \citep[see][for a recent review]{Duc12} even if its origin is not from direct tidal interactions (see Section~5 for a discussion).  Unlike previous examples of tidal dwarfs, AGC~226067 is extremely isolated, $\sim$350~kpc projected distance from the nearest large galaxies.

%a tidal origin for AGC~226067 \citep[see][for a recent review]{Duc12}; unlike previously known tidal dwarfs, AGC~226067 is extremely isolated, $\sim$350~kpc projected distance from the nearest large galaxies; the origin of the system is discussed further in Section~5.  

%This implies a F814W correction factor of 2.1$\pm$0.4 mag from our direct aperture measurement of the diffuse light content, which accounts for the young stellar population below our ACS point source limit.  

%  we assumed a 15 Myr stellar population, with a [Fe/H]=$-$0.6, and integrated the luminosity function \citep[using the appropriate Padova isochrone;][]{Bressan12} below our detection limit
\section{Similar stellar systems in the Virgo Cluster}\label{sec:search}

The optical appearance of AGC~226067 in three-color RGB images, combined with its strong detection in {\it GALEX} imaging, is striking \citep[see][]{Sand15}.  %, and suggests that similar stellar systems may be identified by eye.  
Motivated to understand the uniqueness of AGC~226067 we undertook a search of a 20 deg$^2$ region in the Next Generation Virgo cluster Survey (NGVS) footprint to identify similar systems.%, perhaps as a result of the Virgo cluster environment.  

The NGVS is a $\sim$100 deg$^2$ optical imaging survey conducted with the MegaCam imager on the Canada France Hawaii Telescope (CFHT); details are described in \citet{NGVS}.  %Most of the data discussed here were taken in the $ugiz$ bands, with $r$ band data in select fields.   
The typical point source depth was $g$$\approx$25.9 mag (10$\sigma$).  %This depth is necessary,  as the optical counterpart to AGC~226067 is not visible in SDSS imaging.  
We downloaded fields from the CFHT archive, utilizing the MegaPipe data products \citep{Gwyn08}, in the rectangular region demarcated by fields [1,1] to [-2,-3], using the nomenclature presented in Figure~4 of \citet{NGVS}.  Color RGB images were made with the $u$, $g$ and $i$ band data, and searched by two of us (DJS and DC).  We purposely avoided areas directly adjacent to prominent Virgo cluster galaxies, which often have blue, knotty extensions indicative of extended star formation \citep[e.g.][]{Thilker07} and have a similar appearance as our more isolated examples.  {\it GALEX} \citep{GALEX} imaging was used to support our candidate detections.

Two candidates visually similar to AGC~226067 were identified, and are shown in Figure~\ref{fig:newobjects}.  Each is characterized by an over-density of compact blue sources with  strong GALEX UV emission.  The objects also lack a diffuse reddish galaxy component, exhibited by dwarf galaxies throughout the NGVS imaging.  Aperture photometry \citep{Sand14} is reported for each object in Table~\ref{table:properties}; the objects are of similar brightness as  AGC~226067 in the $NUV,g$ bands to within the uncertainties \citep{Sand15}.
We searched for HI counterparts to our newly identified blue stellar systems using the 40\% ALFALFA catalog \citep{Haynes11}.  No catalogued HI counterparts were found to either source within $\sim$10', and a visual inspection of the ALFALFA data did not turn up any marginal detections.  Assuming  that these objects are similar to AGC~226067, with D=17 Mpc and  velocity width $\sim$50 km s$^{-1}$, we estimate an HI limiting mass of $\sim$2$\times$10$^7$ $M_{\odot}$ for each.  AGC~226067 itself has an $M_{HI}$$\approx$5$\times$10$^7$ $M_{\odot}$ in the ALFALFA catalog, so a similar HI cloud with a slightly higher distance and/or velocity width would not be detectable with the ALFALFA data.  We also note that ALFALFA is blind to emission that spectrally overlaps with the Galactic HI layer at $-$100 $\lesssim$ $v_{lsr}$ (km s$^{-1})$ $\lesssim$ 100 along the line of sight, a velocity range in which they could plausibly be located given that Virgo's Low Velocity cloud has $v_{lsr}$$\approx$0 km s$^{-1}$.

%NGVS-1+2 J122624+142327  NGC4377

%NGVS-1+3 J122631+151026  NGC4419

%{\bf what is their rough luminosity....maybe in GALEX}

%{\bf what is their environment?}

In a future contribution, we will present a full search of the NGVS dataset for similar blue, diffuse objects as well as followup optical/HI observations.  The environment of these diffuse sources will also be investigated, although we note  that NGVS-1+2 J122624+142327 is $\sim$28' ($\sim$130 kpc) from NGC~4377 while NGVS-1+3 J122631+151026  is $\sim$10' ($\sim$50 kpc) from NGC~4419, suggesting their possible association.  Assuming a uniform distribution and given the relative ratio of the entire NGVS and our current search, we can expect to find a sample of $\sim$10-15 systems in total, including AGC~226067 and the newly identified objects.  Taking our inferred star formation rate of the main body of AGC~226067 (1.1$\times$10$^{-3}$ $M_{\odot}$ yr$^{-1}$)  as representative of the population as a whole, it implies an intracluster star formation rate of $\sim$0.01 $M_{\odot}$ yr$^{-1}$ in the Virgo cluster.  The  metallicity of the newly forming stars are comparable to the younger component of intracluster stars \citep{Williams07} and the inferred rate is consistent with the loose upper limit on the intracluster star formation rate found in cluster supernova searches \citep{Graham12}. The detected population may be the tip of the iceberg, as we likely missed less luminous examples, and older stellar populations are more difficult to find because their UV/blue flux fades.  In the past, infalling gas-rich systems were likely more abundant than they are today, thus this process could contribute substantially to the observed population of intracluster stars.

\section{Discussion}\label{sec:discuss}

The properties of AGC~226067 and its immediate environs are intriguing.  In this work, we have shown that the main body has an exclusively young stellar population, with age $\sim$7-50 Myr and [Fe/H]$\approx$$-$0.3, with a $M_{V}$$\approx$$-$11.3 and $M_{star}$$\approx$5$\times$10$^4$ $M_{\odot}$.  %This is consistent with previous work, which noted the presence of a bright ultraviolet GALEX source at the position of AGC~226067, with accompanying HII regions \citep{Bellazzini15UCHVC,Sand15,Beccari16b}.   
There is a secondary group of stars located $\sim$1.5' ($\sim$7.5 kpc) to the Northeast, which is sparse but has a stellar population consistent with the main body of AGC~226067.  %This secondary group of stars is accompanied by a slightly offset, secondary HI cloud \citep{Adams15}.  
The entire stellar system hosts a complex of HII regions that have a nearly uniform metallicity, $\langle$12 + log(O/H)$\rangle$ = 8.37 (for the main body), higher than expected given the stellar luminosity if AGC~226067 were a dwarf galaxy \citep{Beccari16b}.  The HII region velocities of the main and secondary bodies agree with the HI clouds and the overall system displays a weak velocity gradient that spans both bodies \citep{Adams15,Beccari16b}.
%The HII region velocities of the main body ($\langle$$V_r$$\rangle$=$-$153.2$\pm$1.4 km s$^{-1}$) and secondary body ($\langle$$V_r$$\rangle$=$-$126.5$\pm$2.5 km s$^{-1}$)%From our deep HST imaging, there is no evidence for any other extended stellar components.  
Perhaps most intriguingly, there is no old stellar population associated with AGC~226067, as the entire diffuse light budget can be accounted for by the expected faint-end luminosity function of the visible $\sim$7-50 Myr stellar population.  Given the metallicity and lack of old stellar population, AGC~226067 is a strong tidal dwarf candidate, and its extreme isolation makes it a unique example of this class of objects.%We are not aware of such ana tidal%we set a limit of $M_V$$\sim$$-$8 or fainter for this population.  %The presence of 

In understanding the origin of AGC~226067
%The origin of AGC~226067 is ambiguous, but in understanding its origin 
we must also keep in mind that there seem to be other stellar systems of similar appearance in the NGVS dataset (Section~\ref{sec:search}), and perhaps a yet larger population of older, undetectable systems, as well as Virgo HI cloud complexes that have no stellar counterpart at all \citep[e.g.][]{Kent10,Taylor12} -- there is plausibly a continuum of objects that are HI-rich and star-poor \citep[see also the `Almost Dark' objects;][]{Cannon15}.  Recent theoretical effort has sought to understand the physics \citep[e.g.][]{Burkhart16} and possible origins of these clouds \citep[e.g.][]{Taylor16}; here we suggest that AGC~226067 is a stripped remnant of the numerous interactions in the  Virgo cluster environment, most plausibly associated with the M86 subgroup.

%We now discuss two plausible scenarios -- first, that AGC~226067 is a faint, gas-rich dwarf galaxy forming its first generation of stars in the last $\sim$30 Myr and second, that it is a tidal remnant of the numerous interactions in the  Virgo cluster environment, most plausibly associated with the M86 subgroup.

%\subsection{A faint gas-rich dwarf galaxy}

%Based on ground based optical imaging and extensive VLA followup of the HI emission around AGC 226067, \citet{Adams15} concluded that the source was likely a faint, gas-rich dwarf galaxy similar to the SHIELD galaxy sample derived from the ALFALFA catalog \citep{Cannon11}, and that the secondary group of stars and HI gas were likely due to a merger or interaction.  Given the apparent lack of an old stellar population, however, AGC~226067 cannot be considered a typical SHIELD-like galaxy as these systems all have an old red giant branch population at {\it HST} depths \citep{McQuinn14}.

\subsection{A remnant of M86 subgroup interactions}

The removal of gas from cluster galaxies due to the hot intracluster medium (ICM) via ram pressure stripping \citep{Gunn72} is one plausible mechanism that could explain the origin of AGC~226067.  Recent simulations of ram pressure stripping have shown that small amounts of star formation in the stripped tail of ablated gas can be seen up to hundreds of kiloparsecs from the galaxy of origin \citep{Kapferer09,Tonnesen12}.  This star formation is not from stripped molecular clouds of gas from the parent galaxy, but is low density stripped gas that cools and condenses in the turbulent wake of the stripping, a process that takes $\sim$300--750 Myr at the pressure of the Virgo ICM.  Individual star forming knots contain up to $M_{stars}$$\sim$10$^{5-6}$$M_{\odot}$, similar to AGC~226067, and is not expected to contribute significantly to the total intracluster light budget \citep{Kapferer09,Tonnesen12}.  %Previous observations have not been sensitive enough to detect this type of intracluster star formation, which can provide an important test of ram pressure modeling efforts.  %If this mechanism for intracluster star formation at large galactocentric distances is viable, then few observational searches have been sensitive to it, and it is an important verification of ram pressure stripping modeling efforts.
There are examples of intracluster star formation (or implied star formation from H$\alpha$ emission) due to ram pressure stripping in nearby galaxy clusters \citep[e.g.][among others]{Kenney99,Cortese07,Yoshida08,Sun10,Yagi10,Kenney14}, but to our knowledge none have been uncovered at such large projected distances ($D$$\sim$350 kpc) from their point of origin as AGC~226067; systems such as this and those uncovered in this work can provide an important test of ram pressure models.

In Figure~\ref{fig:environ}, we show AGC~226067 in relation to the M86 subgroup of the Virgo cluster, a major component of the `Low Velocity Cloud' of Virgo galaxies with similar low heliocentric velocities as AGC~226067.  AGC~226067 is $\sim$350 kpc in projection from M86.  Overplotted are the Chandra surface brightness contours \citep{Randall08} and H$\alpha$ contours \citep{Kenney08} surrounding both M86 and NGC~4438 which show evidence for a previous interaction with each other \citep[e.g.][]{Kenney08}, and are likely undergoing a ram pressure stripping event as they fall into the larger Virgo cluster together \citep[][among others]{Randall08,Ehlert13}.  The stream of X-ray emitting gas and H$\alpha$ emission point in the general direction of AGC~226067, but the extent of the available datasets is not large enough to probe out to the necessary distances, as marked in Figure~\ref{fig:environ}.  Given the suggestive stream pointing in the direction of AGC~226067, this interpretation of its origin is our preference, although future X-ray and H$\alpha$ mapping will be necessary to confirm this scenario.

 \citet{Beccari16b} suggested that the interacting galaxy pair NGC~4299 ($v_{\odot}$$\approx$+230 km s$^{-1}$) and NGC~4294 ($v_{\odot}$$\approx$+350 km s$^{-1}$) were a plausible origin for AGC~226067 if it were a tidal remnant.  We find this scenario less likely than an M86 subgroup origin, because of the greater projected distance to NGC~4299+NGC~4294 ($\sim$600 kpc), the larger velocity discrepancy between this pair and AGC~226067, and because the HI tidal tail morphology of that pair is pointed to the Southwest, nearly in the opposite direction of AGC~226067 \citep{Chung09}, which is almost due North.
 
% both the greater projected distance to NGC~4299+NGC~4294 ($\sim$600 kpc) and because the HI tidal tail morphology of the pair is pointed to the Southwest, nearly in the opposite direction of AGC~226067 \citep{Chung09}, which is almost due North.

\begin{deluxetable*}{lcccccccccc}
%\rotate
\tablecolumns{2}
\tablecaption{Properties of AGC~226067 \& Other Stellar Systems \label{table:properties}}
\tablehead{
\colhead{Parameter}  & \colhead{Value} \\
%\colhead{} & \colhead{Instrument}
}\\
\startdata
%$m-M$ (mag) & 25.56$\pm$0.16\\
%D (Mpc) &  1.29$\pm$0.1\\
AGC~226067 Main Body \\
\hline
RA (h:m:s) & 12:21:54.04 \\
DEC (d:m:s) & +13:27:35.7 \\
Radial Size (arcsec) & 13.7 \\
Radial Size (kpc) &1.1\\
$M_{V}$ & $-$11.3$\pm$0.3\\
$M_{stars}$  (10$^4$ $M_{\odot}$) & 5.4$\pm$1.3\\
$M_{HI}$ (10$^7$ $M_{\odot}$)\tablenotemark{a} & 1.5\\
$M_{HI}$/$M_{stars}$ & $\approx$280 \\
\hline
AGC~226067 Secondary Body  \\
\hline
RA (h:m:s) & 12:21:55.97 \\
DEC (d:m:s) & +13:28:53.6 \\
Radial Size (arcsec) & 13.0 \\
Radial Size (kpc) & 1.1\\
%$M_{V}$ & \\
$M_{stars}$ (10$^3$ $M_{\odot}$) & 5.5$\pm$4.2\\
\hline
Dw~J122147+132853  \\
\hline
RA (h:m:s) & 12:21:47.87 \\
DEC (d:m:s) & +13:28:54.7 \\
Radial Size (arcsec) & 10.0 \\
Radial Size (kpc) & 0.8 \\
$M_{V}$ & $-$11.4$\pm$0.2\\
\hline
NGVS-1+2 J122624+142327  \\
\hline
RA (h:m:s) & 12:21:24\\
DEC (d:m:s) & +14:23:27 \\
$g$ & 20.1$\pm$0.6\\
$NUV$ & 19.5$\pm$0.3\\
\hline
NGVS-1+3 J122631+151026  \\
\hline
RA (h:m:s) & 12:26:31 \\
DEC (d:m:s) & +15:10:26 \\
$g$ & 20.7$\pm$0.5\\
$NUV$ & 20.8$\pm$0.7
%$M_{V}$ (mag) & $-$10.3$\pm$0.6 \\
%$r_{h}$ (arcsec) & 16.8$\pm$2.4 \\
%$r_{h}$ (pc) & 340$\pm$50 \\
%$\epsilon$ & $<$0.42\tablenotemark{a}\\
\enddata
\tablenotetext{}{The coordinates and radial size for AGC 226067 and  Dw~J122147+132853 are approximate, but correspond to the CMD region used in Figure~\ref{fig:cmds}.  %All physical sizes assume a distance of 17 Mpc.
}
\tablenotetext{a}{From \citet{Adams15}}
%CVnII data: V -- 10 images; 120 sec each; 04/05/2008
%I -- 15 images; 200 sec each; 04/03/2008 & 04/05/2008

\end{deluxetable*}

%\subsection{Implications}

%Intracluster light and star formation.

\section{Conclusions}\label{sec:conclude}

We have presented {\it HST} imaging of the UCHVC AGC~226067, an enigmatic gas-rich stellar system in the Virgo cluster.  The {\it HST} data reveal an exclusively young stellar population of $\sim$7--50 Myr and a [Fe/H]$\sim$$-$0.3, in contrast to normal dwarf galaxy systems which always show some old stellar population ($\gtrsim$5 Gyr) upon close inspection.  Based on these {\it HST} results and other results in the literature on this object (see discussion in Section~\ref{sec:discuss}), there is circumstantial evidence that AGC~226067 is a distant star-forming remnant of the ram pressure stripping event in the M86 subgroup, as recent simulations have predicted \citep{Kapferer09,Tonnesen12}.  

Our initial search for objects with similar optical/UV properties in the NGVS turned up two objects in 20 deg$^2$, but it is likely that we are only sensitive to the youngest and most luminous examples of this emerging class of stellar objects.  Followup  HI, X-ray and H$\alpha$ observations of AGC~226067 and its potential brethren will help elucidate their physical nature.

%Our {\it HST} imaging of AGC~226067 has highlighted the unique properties of this stellar system, which may be either an isolated, gas-rich galaxy just forming stars or a remnant of tidal interactions in the Virgo cluster.  Just as important, our initial search of 20 deg$^2$ of NGVS imaging has yielded two 

%Look for similar stellar systems in all of the NGVS, and study each in detail to determine if this is an emerging class of stellar systems.  

%Sensitive HI, X-ray,  and H alpha observations of AGC~226067 may help solve the mystery of that stellar system.  

%ACKNOWLEDGMENTS.  
\acknowledgments

D.J.S. is supported by NSF
grants AST-1412504 and AST-1517649.  K.S. acknowledges support from the Natural Science and Engineering Research Council of Canada.  E.A.K.A. is supported by TOP1EW.14.105, which is financed by the Netherlands Organisation for Scientific Research (NWO).  E.T. and P.G acknowledge the NSF grants AST-1010038 and AST-1412504.  J.S. acknowledges support from the Packard Foundation.
We thank the  Aspen Center for Physics (NSF Grant \#1066293) for their hospitality during the writing of this paper and the 2016 WoA committee for its support.  Support for program \#13735 was provided by NASA through a grant from the Space Telescope Science Institute, which is operated by the Association of Universities for Research in Astronomy, Inc., under NASA contract NAS 5-26555.

\bibliographystyle{apj}
%\bibliography{mybib}

%\clearpage

%\begin{figure}[ht!]
%\centering
%\includegraphics{HST_UCHVC_2.jpg}
%\caption{A simple caption \label{overflow}}
%\end{figure}

%Your momma's an astronaut

\end{document}